\documentclass[aps,prl,reprint,groupedaddress]{revtex4-1}

\usepackage{graphicx}
\usepackage[caption=false]{subfig}
\usepackage{color}
\usepackage{amsfonts}

\newcommand{\bv}[1]{\mathbf{#1}}
\newcommand{\ket}[1]{\left \vert #1 \right \rangle}
\newcommand{\bra}[1]{\left \langle #1 \right \vert}
\newcommand{\braket}[1]{\left \langle #1 \right \rangle}
\newcommand{\mrm}[1]{\mathrm{#1}}

\begin{document}

\title{Towards hyperpolarization of oil molecules via nitrogen-vacancy centers in diamond}

\author{P. Fern\'{a}ndez-Acebal$^1$, O. Rosolio$^2$, J. Scheuer$^3$, C. M{\"u}ller$^3$, S. M{\"u}ller$^3$, S. Schmitt $^3$, L.P. McGuinness$^3$, I.~Schwarz$^1$, Q. Chen$^1$, A. Retzker$^2$,  B. Naydenov$^3$, F. Jelezko$^3$, M.B. Plenio$^1$}
\affiliation{$^1$ Institut f\"{u}r Theoretische Physik, Albert-Einstein Allee 11, Universitat Ulm, 89069 Ulm, Germany}
\affiliation{$^2$ Racah Institute of Physics, The Hebrew University of Jerusalem, Jerusalem, 91904 Givat Ram, Israel}
\affiliation{$^3$ Institute for Quantum Optics, Ulm University, Albert-Einstein-Allee 11, Ulm 89081, Germany}

\date{\today}

\begin{abstract}
Efficient polarization of organic molecules is of extraordinary relevance when performing nuclear magnetic resonance (NMR) and imaging. 
Commercially available routes to dynamical nuclear polarization (DNP) work at extremely low-temperatures, thus bringing the molecules out of their ambient thermal conditions and relying on the solidification of organic samples.
In this work we investigate polarization transfer from optically-pumped nitrogen vacancy centers in diamond to external molecules at room temperature.
This polarization transfer is described by both an extensive analytical analysis and numerical simulations based on spin bath bosonization and is supported by experimental data in excellent agreement.
These results set the route to hyperpolarization of diffusive molecules in different scenarios and consequently, due to increased signal, to high-resolution NMR.
\end{abstract}

\maketitle

\textit{Introduction.---} Nuclear magnetic resonance (NMR) is a fundamental tool in the biomedical sciences \cite{PhysRev.70.474,PhysRev.69.37}. As the sensitivity of NMR is proportional to the sample polarization, hyperpolarized samples, where the population difference between nuclear spins exceeds significantly its thermal value, are desirable for achieving a higher NMR signal. One of the promising methods achieving such a hyperpolarization is dynamic nuclear polarization (DNP) in which a polarized electron spin transfers its polarization to a nuclear bath via dipolar coupling. While currently commercially available techniques require cryogenic temperatures to polarize the electron spins, we take a different route based on optical polarization of nitrogen vacancy (NV) centers in diamonds at room-temperature.

NV centers are negatively charged paramagnetic defects in diamond with unpaired electronic spin triplet in their ground state \cite{wu2016diamond}. These color centers can be significantly polarized (exceeding $92\%$) by optical pumping without the need for low temperatures nor high magnetic fields \cite{waldherr2011dark}. This polarization process can be achieved in less than a microsecond while the NV center relaxation time can be in the order of milliseconds. 
Furthermore, the highly polarized NV center electron spins can be brought into resonance with adjacent nuclear spins, making them great candidates for DNP.

Special interest over the last years has been paid to hyperpolarization of molecules in solution \cite{PhysRevB.92.184420,PhysRevB.93.060408,abrams2014dynamic,ardenkjaer2003increase,joo2006situ}. Due to the molecular motion and short correlation times in fluids, the anisotropic interaction between the electron and the target spins averages out. Consequently, cross-relaxation mechanisms are commonly used in polarization of liquids at ambient conditions since common methods developed for DNP of stationary spins such as the solid effect \cite{vanBentum2016126,Wiśniewski201630} are less efficient.

In this work, we demonstrate that polarization loss from a single shallow implanted NV center can be understood as the result of polarization transfer to $^1H$ nuclei in surrounding oil molecules. As the NV center is several nanometers away from the diffusing molecules and oil exhibits high viscosity at room-temperature, the nuclei diffuse in and out of the NV center interaction region in a time scale comparable with the interaction strength. This results in an increase of the polarization rate enabling resonant transfer~\cite{PhysRevB.93.060408}. We use the Hartmann-Hahn double resonance (HHDR) scheme \cite{hartmann1962nuclear}, where the electron spin is driven with a Rabi frequency that matches the Larmor frequency of diffusive nuclei. Consequently, the interaction between the NV center and the $^1H$ nuclei is strengthened, while the effects of noise sources on the NV center are weakened. Thus, HHDR scheme serves as a continuous dynamical decoupling (CDD) protocol \cite{cai2013,PhysRevA.58.2733,PhysRevLett.109.070502,facchi2004unification,fanchini2007continuously,gordon2008optimal,Timoney2011,webster2013simple,aharon2013general} allowing us to polarize the specified nuclear species efficiently. In fact, our method is limited by the relaxation time in the rotating frame, $T_{1 \rho}$, (finding its origin in $T_2$ processes in the non-rotating frame), in contrast to recently proposed microwave-free protocols that are limited by the much shorter dephasing time $T_2^*$ \cite{wood2017microwave}.

This Letter is organized as follows: First the Hamiltonian describing the interaction between one electronic (NV) spin and a bath of $N$ diffusive nuclear spins is presented. We then derive an analytical solution for the polarization loss of the NV center. In order to compare the theoretical prediction to a robust numerical simulation and due to the fact that large spin baths can not be fully simulated, we use the Holstein-Primakoff approximation (HPA) \cite{PhysRev.58.1098}, which considers spins as bosons, allowing us to perform an efficient numerical implementation. Finally, we validate our theoretical findings, carried out with no free parameters, by performing experiments with two different shallow NV centers coupled to oil molecules. Theory and experiment show an excellent agreement and therefore, support that polarization loss of the NV center is best explained by polarization transfer to oil at room-temperature.

\begin{figure}
	\includegraphics[width=0.8\columnwidth]{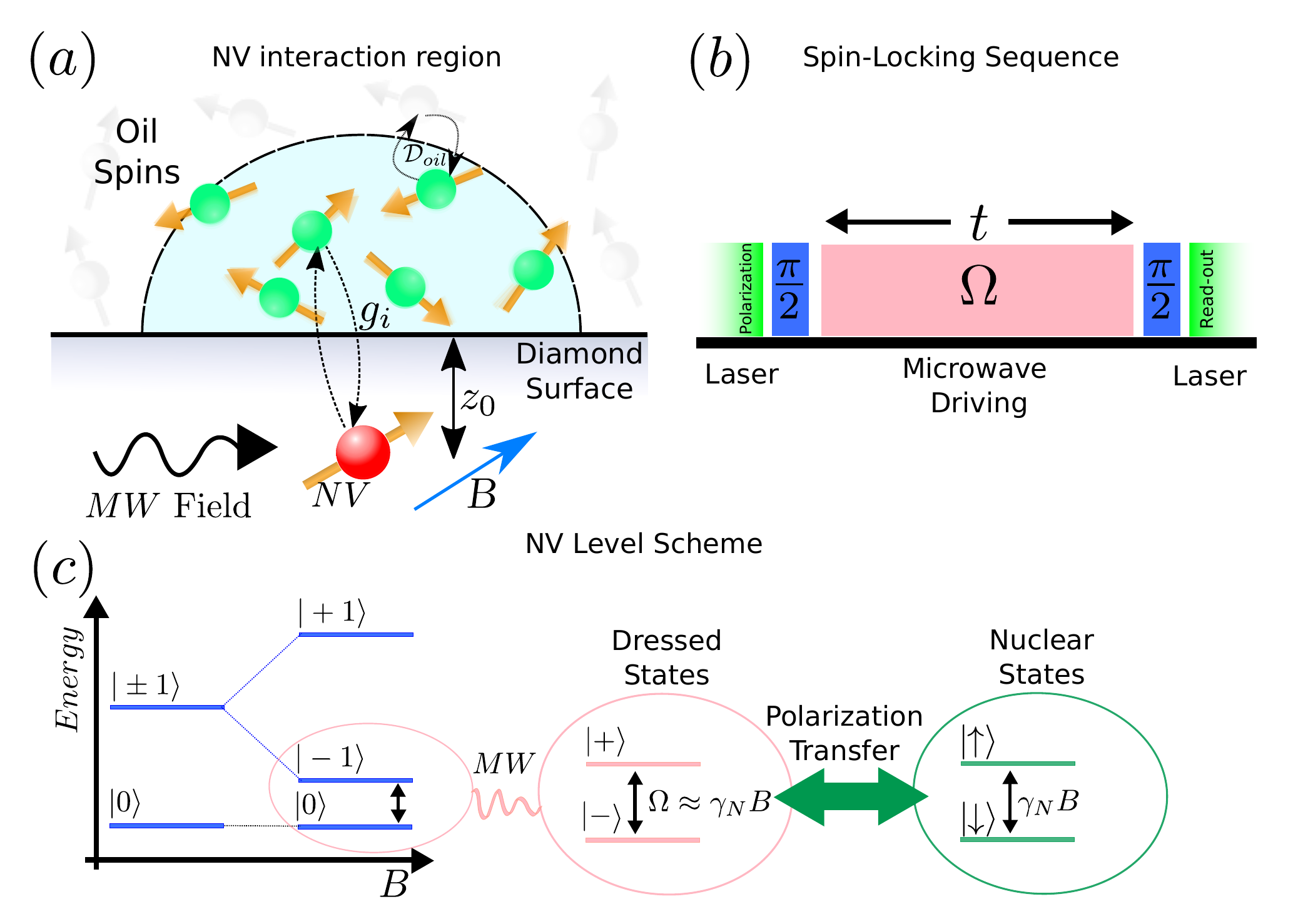}
	\caption{ Schematic set-up for hyperpolarization of oil molecules with shallow NV centers in bulk diamond and important parameters. (a) The NV center is located at a distance $z_0$ beneath the diamond surface. The quantization axis of the NV center coincides with the $[1,1,1]$ direction inside the diamond lattice. The oil is deposited on top of the diamond. The $^1H$ spins conforming the oil move stochastically with a diffusion coefficient ${\cal{D}}_{oil}$, by means of this motion they diffuse in and out the detection volume (blue semi-sphere). Only the $N$ spins inside the volume (in green) are considered, while the interaction with outer spins (in grey) is neglected. (b) Spin-Locking sequence applied to measure the NV center polarization loss. The NV center is initially polarized and read-out using green-laser. During the microwave driving time, $t$, the NV center interacts with the nuclear spins. (c) Energy levels of the NV center. In the absence of magnetic field the NV center is a spin$-1$ system with projections $m_s = 0, \pm 1$. Degeneracy between $m_s=\pm 1$ levels is lifted by applying an external magnetic filed, $B$, parallel to the NV center axis. The transition between $m_s= 0$ and $m_s=-1$ is driven with a microwave field inducing a Rabi frequency $ \Omega$ which matches the nuclear Larmor frequency $\omega_N = \gamma_N B$ permitting polarization transfer. 
	}
	\label{fig:Fig1}
\end{figure}

\textit{System.---} We use a shallow NV center in bulk diamond located at a distance $z_0$ underneath the diamond surface. Immersion oil containing the $^1H$ nuclei is deposited onto the surface. The set-up is depicted in Fig(\ref{fig:Fig1}). At room temperature the oil molecules diffuse with diffusion coefficient ${\cal{D}}_{oil}$, which results in a finite correlation time, $\tau_c$, for the electron-nuclear spin interaction \cite{SupplementalMaterial,staudacher2015probing,PhysRevApplied.4.024004}.

In solid samples at HHDR, flip-flop transitions (polarization transfer) between electron and nuclear spins predominate over flip-flip processes (depolarization), which are energetically forbidden, thus generating a net polarization transfer at weak coupling. For liquid samples, however, the transfer efficiency is highly determined by a parameter $\chi \equiv \omega_N \tau_c$ \cite{PhysRevB.93.060408,SupplementalMaterial}, with $\omega_N$ the Larmor frequency of the nuclei in diffusion. For $\chi \ll 1$, as is typical for strongly diffusing molecules, such as water, effective interaction time is shorter than $\omega_N^{-1}$, in this regime the imbalance between flip-flip and flip-flop transitions is suppressed, even on resonance, so no net polarization is achieved. However, as the NV center is several nm apart from the diffusing molecules, and oil possesses a high viscosity at room-temperature (${\cal{D}}_{oil} \approx 0.5 \, \mrm{nm}^2 \mu \mrm{s}^{-1}$), the resulting correlation time is large, making $\chi \gg 1$, for shallow NV centers ($z_0 \approx 3-5 \, \mrm{nm}$). Therefore, once the HHDR condition is achieved, flip-flip transitions are suppressed by fast rotations so that in the presence of flip-flop transitions an efficient polarization transfer from the electronic spin to the $^1H$ nuclei is expected.	

\textit{Spin model.---}We consider a single NV center, electronic spin$-1$, interacting with $N$ diffusive nuclear spins$-1/2$. In the secular approximation \cite{london2013detecting}, this interaction is described by $H_{int} = I_Z^{NV} \sum_i \bv{A}^i(t) \bv{I}_i$, where $\bv{I}^{NV}$ is the spin$-1$ operator of the NV center, with projections $ m_s = 0, \pm 1$, $ \bv{I}_i$ is the spin operator of the $i^{th}$ nucleus and $\bv{A}^i(t)$ is the hyperfine vector between them \cite{SupplementalMaterial}. The NV center is driven with a microwave field with Rabi frequency $\Omega$ resonant with the $m_s=0 \rightarrow m_s=-1$ transition, creating an effective two-level system. In the presence of an external magnetic field $\bv{B}$, the nuclear Larmor frequency is $\omega_N = \gamma_N \vert \bv{B} \vert$, with $\gamma_N$ the nuclear gyromagnetic ratio. The magnetic field generated by the driven NV center perturbs the external field acting on the nuclear spins, hence $\omega_N$ fluctuates due to this effect. Polarization transfer occurs at HHDR condition, $ \Omega = \omega_N$. In the dressed-state basis, $\ket{\pm} = \frac{1}{\sqrt{2}} \left( \ket{0} \pm \ket{-1} \right) $ for the NV center and assuming $\tau_c \omega_N \gg 1$ in our setup, we can assume the rotating wave approximation to find the effective Hamiltonian ($\hbar = 1$)

\begin{equation}
\label{eq:TotalHamiltonian}
H= \Omega S_z + \sum_{i=1}^N \omega_N I_z^i + \sum_{i=1}^N g_i(t) S_+ I_-^i + g_i^*(t) S_-I_+^i,
\end{equation}
with $S_z = \frac{1}{2} \left( \ket{+}\bra{+} - \ket{-}\bra{-} \right)$ and the coupling strength $g_i(\bv{x}_i(t)) \equiv g_i(t) = \frac{1}{4} \left( A_x^{i}(t) + {\mrm{i}} A_y^{i}(t) \right)$, where $\bv{x}_i(t)$ is the relative position between the NV and the $i^{th}$ nucleus, and $A_{\alpha}^i (t)$ are the different components of the hyperfine vector $\bv{A}^i (t)$. Note that, $g_i(t)$ is a stochastic variable with certain correlation time $\tau_c$.
 
Indeed, over time intervals larger than $\tau_c$, the system state may be expressed as $ \braket{\rho}(t) = \braket{\rho}_{NV} (t) \otimes \rho_B$, with $\rho_B = \bigotimes_{i=1}^N \frac{1}{2} \mathbb{I} $ the thermal state of the nuclei, $\braket{\rho}$ the system density matrix averaged over all stochastic trajectories of $g_i(t)$, and $\braket{\rho}_{NV}$ the NV center density matrix. Within this description, correlations among spins are neglected, which allows us to obtain for the average NV center population, $\braket{n} = \frac{1}{2} + \mrm{Tr} \left( S_z \braket{\rho}_{NV} \right) $, the dynamical equation
 
 \begin{equation}
 \label{eq:PopulationEquation}
 \dot{\braket{n}}+ \frac{1}{4} N \gamma(t) \braket{n} = \frac{1}{4} N \gamma(t) \braket{n}_B, 
 \end{equation} 
 where we define $ \gamma(t) = \int_0^t \braket{\xi_\alpha^i (t') \xi_\alpha^i (0)} d t'$, $\xi_\alpha^i(t) = A_\alpha^i(t)-\braket{A_\alpha^i(t)}$, and $ \braket{n}_B$ is the average nuclear population, which is $\braket{n}_B = 1/2$ for thermal bath. Eq.(\ref{eq:PopulationEquation}) reflects a net incoherent polarization transfer towards the nuclear reservoir.

The description of $A_\alpha^i(t)$ as a stochastic variable is not trivial. Even if the nuclei undergo thermal random motion, which is Gaussian and thus easy to characterize, the intricate dependence of $A_\alpha^i(t)$ on the relative position between NV center and nuclei makes $A_\alpha^i(t)$ itself not a Gaussian variable. Consequently, the derivation of a close analytic solution for $\braket{A_\alpha^i}$ or $\gamma$ is not straightforward. Instead, a full analytical description of the polarization evolution may be derived when assuming $A_\alpha^i(t)$ is such that: i) Its higher order cumulants are negligible compared to its mean and variance. ii) The correlation decays exponentially, then $ \gamma(t) = \sigma^2 \tau_c \left( 1-\exp(-t/\tau_c) \right)$, with $\sigma^2$ the variance of $A_\alpha^i(t)$. Note we have dropped $\alpha$ and $i$ since for homogeneous diffusion all the nuclei have equal average properties. These two assumptions have been numerically tested, see \cite{SupplementalMaterial}. With these considerations the solution of Eq.~(\ref{eq:PopulationEquation}) is

\begin{equation}
\label{eq:PopulationSolution}
\braket{n} = \frac{1}{2} + \frac{1}{2} \exp \left( \frac{1}{4} N \tau_c^2 \sigma^2 \left( 1 - t/\tau_c - e^{-t/\tau_c} \right) \right).
\end{equation}
For $t \ll \tau_c$,  $\braket{n}$ shows a Gaussian decay with a polarization rate $ \sqrt{\frac{1}{8}  N \sigma^2}$. On the other hand, for $t \gg \tau_c$ we find $\braket{n}$ exhibits an exponential behavior with rate $  \frac{1}{4} N \sigma^2 \tau_c$. Eqs.(\ref{eq:PopulationEquation}-\ref{eq:PopulationSolution}) are only valid for times such that $ N \braket{{\xi_\alpha^i}^3} \tau_c^2 \ll t^{-1} $ and $ N \braket{g}^2 \tau_c \ll t^{-1} $ \cite{SupplementalMaterial} (note that for non-Gaussian variables the third moment does not vanish). The derived equations describe the polarization dynamics of the system and do not include $T_{1 \rho}$ relaxation. For a discussion of the latter see \cite{SupplementalMaterial}.

\textit{Boson  Model.---}Exact computational simulations of large spin systems are not possible, thus different approximations have been taken to model the behavior of large spin baths \cite{PhysRevB.75.155324}. We perform simulations based on Gaussian states to compute the dynamical evolution of a system ruled by Eq.(\ref{eq:TotalHamiltonian}). Specifically we make use of Holstein-Primakoff approximation (HPA) \cite{PhysRev.58.1098} representing a polarized spin as a boson in its ground state. In the lowest order of the approximation, the spin operators are substituted by bosonic operators, $ S_- \rightarrow a$. This approximation holds for highly-polarized spins while for spins in its thermal state it offers a lower bound to the polarization dynamics \cite{PhysRevB.75.155324}. The bosonic Hamiltonian is then obtained from Eq.(\ref{eq:TotalHamiltonian}) as

\begin{equation}
	\label{eq:BosonHamiltonian}
	H= \Omega a^\dagger a + \sum_{i=1}^N \omega_N {b_i}^\dagger b_i + \sum_{i=1}^N g_i a^\dagger b_i + g_i^* a {b_i}^\dagger,
\end{equation}
which is quadratic in the operators $\{ a,a^{\dagger},b_i,b_i^{\dagger} \}$, allowing an efficient numerical simulation describing the dynamical evolution via the covariance matrix \cite{plenio2004dynamics,eisert2003introduction,SupplementalMaterial}.

\begin{figure}[htp]
		\includegraphics[clip,width=0.8\columnwidth]{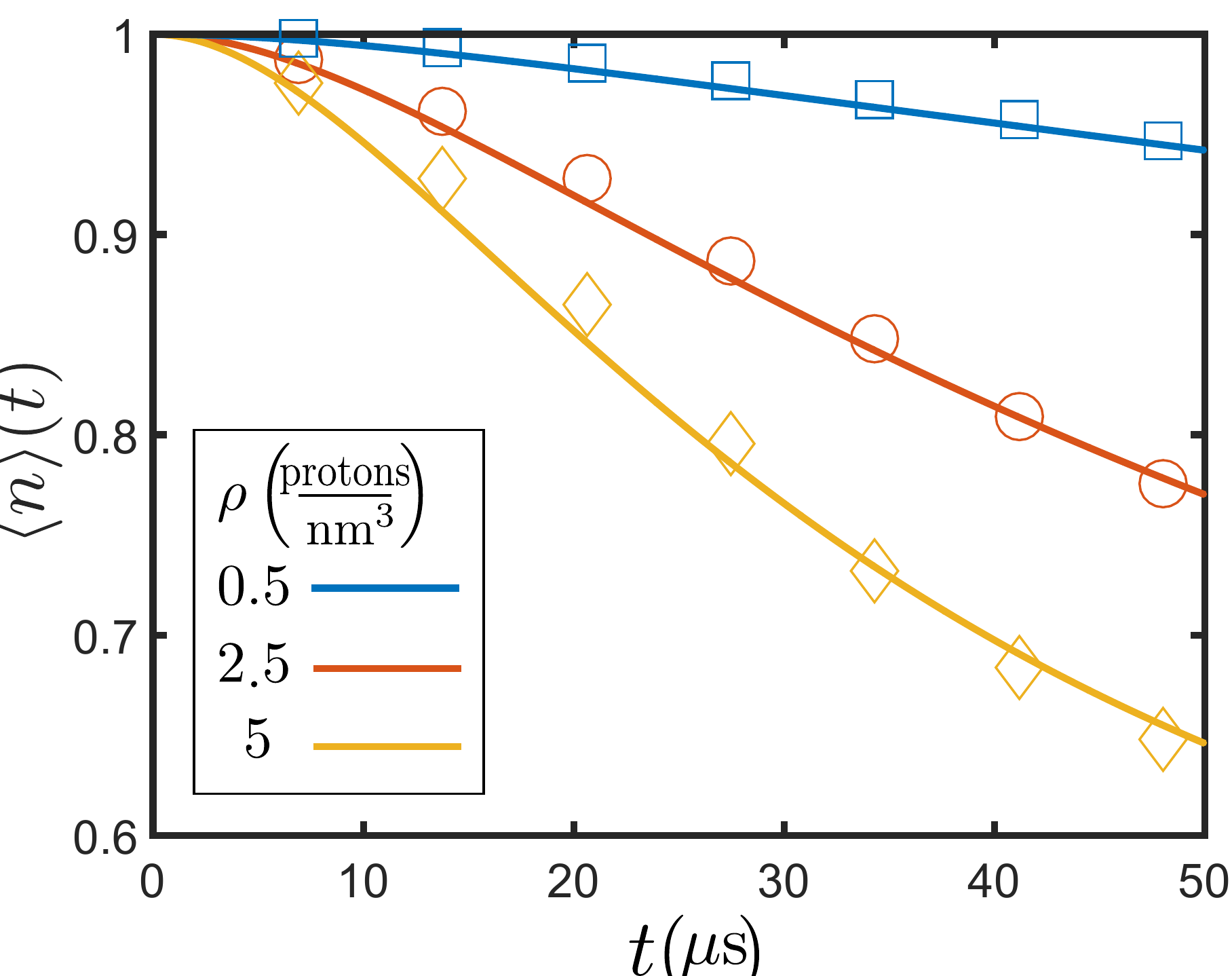}
	\caption{Comparison between theoretical predictions, Eq. (\ref{eq:PopulationEquation}) for a spin system (lines), and full numerical simulations using the bosonic approximation, Eq.(\ref{eq:BosonHamiltonian}) (markers) for different sample densities and fixed ${\cal{D}}_{oil}$ and $z_0$. For the theory curve, the values $\sigma^2$ and $\tau_c$ are calculated via direct numerical integration.} 
	\label{fig:FigureNumerics}
\end{figure}

\begin{figure*}[htp]
	
	\includegraphics[clip,width=1.6\columnwidth]{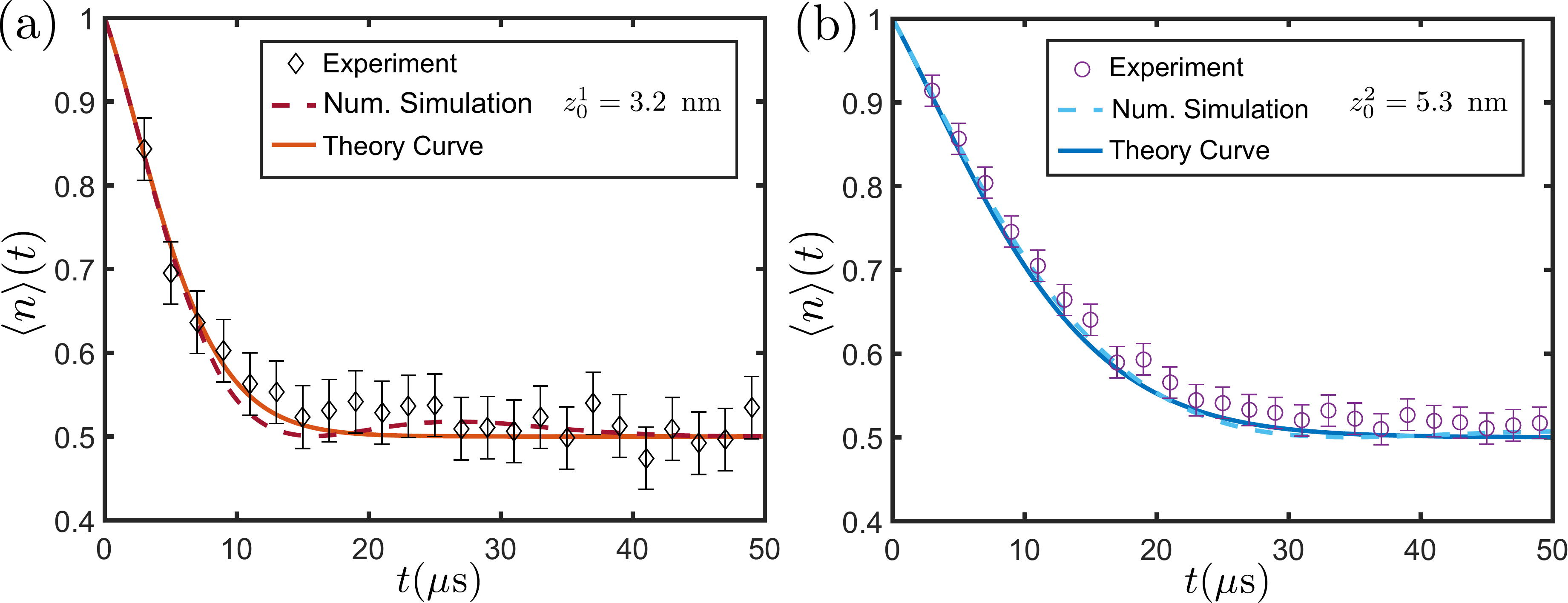}%
	
	\caption{ Measured polarization loss from two different NV centers compared with numerical and theoretical predictions. The qualitative behavior is well predicted by both theory and full numerical simulation. The polarization interchanged dictated by Eq.(\ref{eq:PopulationSolution}) happens in the same time scale as relaxation processes characterized by $T_{1_\rho}$. As a result, an initial decay determined by $-t/T_{1 \rho}$ is followed by a exponential tail, with rate $\left(  \frac{1}{4} N \tau_c \sigma^2 + \frac{1}{T_{1\rho}} \right)$. (a) Results for the shallower NV center, $z_0^1 = 3.2 \, \mrm{nm}$. After an initial decay, at $t\approx \, 25 \, \mu \mrm{s}$ numerical simulations reveal a coherent interchange of polarization between NV center and slowly moving nuclei near the surface. This feature is not captured in our theoretical prediction. (b) Result for $z_0^2 = 5.3 \, \mrm{nm}$. For deeper NV center, the polarization is better described by Eq.(\ref{eq:PopulationSolution}) with no traces of coherent transfer. \{In the calculations, have been used ${\cal{D}}_{oil} = 0.46 \, {\mrm{nm}}^2 \mu {\mrm{s}}^{-1}$, $\rho_{oil} = 50 \, {\mrm{spins}}/{\mrm{nm}}^{3}$, which are directly estimated from the experimental conditions, together with the experimentally measured relaxation times $T_{1\rho}^1 = 11 \, \mu \mrm{s}$ and $T_{1\rho}^2 = 17 \, \mu \mrm{s}$ \cite{SupplementalMaterial}.\} 
		}
	\label{fig:FigureResults}
\end{figure*}

The full numerical simulation is performed considering $N$ independently moving nuclei in a finite box with periodic boundary conditions. The box represents the detection volume of the NV center and its length is proportional to the NV center depth \cite{Staudacher561,staudacher2015probing}. When a nucleus crosses the box walls, it is substituted by a nucleus from the reservoir, thus losing its correlation with the NV center and the rest of spins. Each particle describes a $3-$dimensional Brownian motion \cite{gardiner1985handbook}. Due to the molecular motion, internuclear coupling among nuclei is averaged out. Therefore, specific molecular structure is not relevant in the main dynamics. Rotations and vibrations of the molecule are not considered in the simulations. Results showing the agreement between the solution of Eq.(\ref{eq:PopulationEquation}) and a full numerical simulation are depicted in Fig.(\ref{fig:FigureNumerics}). We remark that when using HPA, correlations among spins build up during the time the nuclei are diffusing inside the box. As a consequence, the presence of slowly moving nuclei in the diamond vicinity may cause coherent polarization transfer, see Fig.(\ref{fig:FigureResults}). This feature is not captured in  our theoretical description, Eq.(\ref{eq:PopulationEquation}).

The polarization rate depends solely on $N$, $\sigma^2$ and $\tau_c$ that are extensive parameters which in turn depend on intensive quantities, namely, $\rho_{oil}$, the proton density in oil, ${\cal{D}}_{oil}$ and $z_0$. Once $ \lbrace \rho_{oil}, {\cal{D}}_{oil}, z_0 \rbrace $ are fixed, the polarization curve is obtained directly by a numerical simulation, while the extensive parameters may be calculated via a numerical integration \cite{SupplementalMaterial} and then used as input for Eq. (\ref{eq:PopulationSolution}). Therefore, our model is without free fitting parameters.

\textit{Experimental Implementation.---} For an experimental verification of our model, the polarization loss from two different NV centers was measured. It is observed via the NV's fluorescence after a spin-locking sequence matching the HHDR with the hydrogen nuclear spins. First, a single NV is optically polarized using a $ 532 \, \mrm{nm}$ laser. Followed by a MW $\pi/2$ pulse, that rotates the NV's electron spin phase-dependent to the $\ket{+}$ or $\ket{-}$ state in an alternating manner. Then, a suitable microwave field is applied in order to fulfill the HHDR condition. During this period the NV transfers its high polarization to the environment. A final $\pi/2$ pulse projects the NV center's spin state back to the z-axis, where it can be read out optically. Depending on the occurrence of a spin flip a bright or dark fluorescence signal is observed. The measuring protocol is afterwards repeated for the second NV center.

The $^{12}C$ enriched diamond sample was grown by chemical vapour deposition with low concentration of impurities ($^{12}C >99.999 \% $ ; nitrogen isotopes $\approx 5 \, \mrm{ppb}$ ). Nitrogen ions were implanted with a low energy and dose ($2.5 \, \mrm{keV}$ and $10^8$ ions/cm$^2$) such that experiments can be performed with shallow NV centers ( here: $z_0^1 = 3.2 \pm 0.2 \, \mrm{nm}$ and $ z_0^2 = 5.3 \pm 0.1 \, \mrm{nm}$ \cite{PhysRevB.93.045425}), ensuring strong coupling with near-surface nuclei. Immersion oil (Fluka Analytical 10976) is deposited on top of the diamond. A magnetic field of $ 660 \, \mrm{G}$ is applied parallel to the NV center quantization axis, resulting in a $^1{H}$ Larmor frequency $\omega_N/(2  \pi) = 2.8 \, \mrm{MHz}$, where the effective field generated by the NV center can be neglected. The induced Rabi frequency during the spin locking pulse is adjusted to be resonant with $\omega_N$, to allow electron spin - nuclear spin flip-flop processes. The experiments were performed on a home-build confocal microscope controlled with the Qudi software suite \cite{binder2017qudi} .

Fig.(\ref{fig:FigureResults}) shows the comparison between the measured polarization loss, $\braket{n}$, the full numerical simulation based on Gaussian states and the theoretical prediction from Eq.(\ref{eq:PopulationSolution}). In both cases, we observe polarization loss in a time scale of the order of $\mu s$. Off-resonant measurements reveal that environmental magnetic noise affecting the NV center manifests in similar time scales. These effects are included in our predictions by adding a relaxation term to the master equation proportional to $T_{1\rho}$ \cite{slichter1990principles,SupplementalMaterial}.

\textit{Discussion.---}The experimental data is well reproduced given the known ${\cal{D}}_{oil}$, NV depth, the proton density in oil at room-temperature $\rho_{oil}=50 \, \mrm{protons/nm}^3$, which is a typical value used for organic samples \cite{Staudacher561,meriles2010imaging} and the measured relaxation times, ($T_{1\rho}^1 = 11 \, \mu \mrm{s}$, $T_{1\rho}^2 = 17 \, \mu \mrm{s}$) \cite{SupplementalMaterial}. For this density, the number of hydrogen spins in an interaction volume of $20^3 \, \mrm{nm}^3$, which is taken as the volume of the simulation box, is $N\approx 4 \cdot 10^5$ (See \cite{SupplementalMaterial} for details in the simulation technique). Given the depths of the used NV centers ($z_0^1= 3.2 \, {\mrm{nm}}$, $z_0^2= 5.3 \, {\mrm{nm}}$) and the slow diffusion of the oil molecules (${\cal{D}}_{oil} = 0.5 \, \mrm{nm}^2 \mu \mrm{s}^{-1}$), the correlation time was numerically calculated \cite{SupplementalMaterial}, obtaining $\tau_c^1 = 10 \, \mu \mrm{s}$ and $\tau_c^2 = 25 \, \mu \mrm{s}$ respectively. As stated, for $\chi \gg 1$ flip-flop dominates over flip-flip rate \cite{SupplementalMaterial}, hence the latter does not produce a significant effect.

Since molecular motion effectively broadens the nuclear lines, our scheme has the advantage of relatively high robustness to frequency detuning (e.g. from fluctuations in the magnetic field) and Rabi frequency errors. Also, we have exclusively considered particles in free-diffusion, which is an accurate description for molecules in bulk but may not describe molecular motion at the oil-diamond interphase  correctly. Nonetheless, short-time dynamics, most relevant for hyperpolarization protocols, are well described solely by diffusive particles, and near-surface traces will be seen only at longer times.

The good agreement between our model and the experimental data supports the hypothesis that the NV polarization loss mainly occurs due to polarization transfer to the $^1$H nuclei and $T_{1 \rho}$ relaxation. Therefore, the amount of transfered polarization is just a fraction of the total, $\alpha=\frac {1/ \tau_p}{1/\tau_p + 1/ T_{1 \rho} }$, where $\tau_p^{-1} \equiv \frac{1}{4} N \tau_c \sigma^2$ is the polarization rate. In our set-up we obtain $\alpha \sim 80 \% $ for both NV centers, indicating that transfer is efficient. Still, the average polarization gain per nuclei around the NV is $ \frac{1}{2} \frac{1}{N} \alpha$, which is very small for large $N$. Thus, significant nuclear polarization will only be achieved when the proposed protocol is repeated many times. That is, the NV center must be periodically reinitialized after a time $\tau_p$. 

Assuming a diamond sample with high density of NV centers, a thin layer of oil of few microns deposited onto its surface and typical nuclear relaxation time $T_{1n} \approx 1 \, \mrm{s}$ \cite{Fullerton1982209,Shapiro20111184}, the maximum achievable polarization per nuclei may be estimated as $P_n \approx 10^{-3}$ \cite{SupplementalMaterial}, which exceeds by several orders of magnitude the thermal nuclear polarization at this temperature and field $P_n^{Th} \approx 10^{-7}$.   

Besides, writing the dependence of the polarization rate explicitly as a function of the diffusion coefficient, proton density and NV depth, one obtains that $\frac{1}{\tau_p} \propto \frac{ \rho_{^1 {\mrm{H}} } }{z_0 \cal{D}}$ \cite{SupplementalMaterial}. Thus, making it easy to estimate the polarization rate for different solvents or NV samples.

\textit{Conclusion.---}In summary, we present a theoretical description of efficient polarization transfer from a shallow NV center to diffusive organic molecules above the diamond surface validated by experiments with an excellent agreement. The experimental results are explained qualitatively and quantitatively by both theoretical work and numerical simulations. We remark, our approach is easily extended to other scenarios such as polarization of macromolecules in low diffusive environments or polarization schemes using nanodiamonds. In fact, our model does not depend on any free parameters and thus the polarization rate depends solely on the NV depth and the diffusion properties of the solvent. Also, the dynamical behavior at the liquid-solid interface may be examined with our description. Due to the fast diffusion of the nuclei, the direct detection of polarization remains a challenge that needs to be addressed in future work.

\begin{acknowledgments}
 This work was supported by an Alexander von Humboldt Professorship, the ERC Synergy grant BioQ and the EU projects HYPERDIAMOND and DIADEMS.
\end{acknowledgments}

\bibliographystyle{unsrt}

\begin{thebibliography}{10}
	
	\bibitem{PhysRev.70.474}
	F.~Bloch, W.~W. Hansen, and M.~Packard.
	\newblock The nuclear induction experiment.
	\newblock {\em Phys. Rev.}, 70:474--485, 1946.
	
	\bibitem{PhysRev.69.37}
	E.~M. Purcell, H.~C. Torrey, and R.~V. Pound.
	\newblock Resonance absorption by nuclear magnetic moments in a solid.
	\newblock {\em Phys. Rev.}, 69:37--38, 1946.
	
	\bibitem{wu2016diamond}
	Y.~Wu, F.~Jelezko, M.~B. Plenio, and T.~Weil.
	\newblock Diamond quantum devices in biology.
	\newblock {\em Angew. Chem., Int. Ed.}, 55(23):6586--6598, 2016.
	
	\bibitem{waldherr2011dark}
	G.~Waldherr, J.~Beck, M.~Steiner, P.~Neumann, A.~Gali, T.~Frauenheim,
	F.~Jelezko, and J.~Wrachtrup.
	\newblock {Dark states of single nitrogen-vacancy centers in diamond unraveled
		by single shot NMR}.
	\newblock {\em Phys. Rev. Lett.}, 106(15):157601, 2011.
	
	\bibitem{PhysRevB.92.184420}
	Q.~Chen, I.~Schwarz, F.~Jelezko, A.~Retzker, and M.~B. Plenio.
	\newblock Optical hyperpolarization of $^{13}\mathrm{C}$ nuclear spins in
	nanodiamond ensembles.
	\newblock {\em Phys. Rev. B}, 92:184420, 2015.
	
	\bibitem{PhysRevB.93.060408}
	Q.~Chen, I.~Schwarz, F.~Jelezko, A.~Retzker, and M.~B. Plenio.
	\newblock Resonance-inclined optical nuclear spin polarization of liquids in
	diamond structures.
	\newblock {\em Phys. Rev. B}, 93:060408, 2016.
	
	\bibitem{abrams2014dynamic}
	D.~Abrams, M.~E. Trusheim, D.~R. Englund, M.~D. Shattuck, and C.~A. Meriles.
	\newblock Dynamic nuclear spin polarization of liquids and gases in contact
	with nanostructured diamond.
	\newblock {\em Nano Lett.}, 14:2471--2478, 2014.
	
	\bibitem{ardenkjaer2003increase}
	J.~H. Ardenkj{\ae}r-Larsen, B.~Fridlund, A.~Gram, G.~Hansson, L.~Hansson, M.~H.
	Lerche, R.~Servin, M.~Thaning, and K.~Golman.
	\newblock Increase in signal-to-noise ratio of $>$ 10,000 times in liquid-state
	{NMR}.
	\newblock {\em Proc. Natl. Acad. Sci. U.S.A.}, 100(18):10158--10163, 2003.
	
	\bibitem{joo2006situ}
	C.~Joo, K.~Hu, J.~A. Bryant, and R.~G. Griffin.
	\newblock In situ temperature jump high-frequency dynamic nuclear polarization
	experiments: Enhanced sensitivity in liquid-state {NMR} spectroscopy.
	\newblock {\em J. Am. Chem. Soc.}, 128(29):9428--9432, 2006.
	
	\bibitem{vanBentum2016126}
	P.J.M. van Bentum, M.~Sharma, S.G.J. van Meerten, and A.P.M. Kentgens.
	\newblock Solid effect $\mathrm{DNP}$ in a rapid-melt setup.
	\newblock {\em J. Magn. Reson.}, 263:126--135, 2016.
	
	\bibitem{Wiśniewski201630}
	D.~Wi{\'s}niewski, A.~Karabanov, I.~Lesanovsky, and W.~K{\"o}ckenberger.
	\newblock Solid effect $\mathrm{DNP}$ polarization dynamics in a system of many
	spins.
	\newblock {\em J. of Magn. Reson.}, 264:30--38, 2016.
	
	\bibitem{hartmann1962nuclear}
	S.~R. Hartmann and E.~L. Hahn.
	\newblock Nuclear double resonance in the rotating frame.
	\newblock {\em Phys. Rev.}, 128:2042, 1962.
	
	\bibitem{cai2013}
	J.~Cai, F.~Jelezko, M.B. Plenio, and A.~Retzker.
	\newblock Diamond-based single-molecule magnetic resonance spectroscopy.
	\newblock {\em New J. Phy.}, 15:013020, 2013.
	
	\bibitem{PhysRevA.58.2733}
	L.~Viola and S.~Lloyd.
	\newblock Dynamical suppression of decoherence in two-state quantum systems.
	\newblock {\em Phys. Rev. A}, 58(4):2733--2744, 1998.
	
	\bibitem{PhysRevLett.109.070502}
	X.~Xu, Z.~Wang, C.~Duan, P.~Huang, P.~Wang, Y.~Wang, N.~Xu, X.~Kong, F.~Shi,
	X.~Rong, and J.~Du.
	\newblock Coherence-protected quantum gate by continuous dynamical decoupling
	in diamond.
	\newblock {\em Phys. Rev. Lett.}, 109(7):070502, 2012.
	
	\bibitem{facchi2004unification}
	P.~Facchi, D.A. Lidar, and S.~Pascazio.
	\newblock {Unification of dynamical decoupling and the quantum Zeno effect}.
	\newblock {\em Phys. Rev. A}, 69(3):032314, 2004.
	
	\bibitem{fanchini2007continuously}
	F.F. Fanchini, J.E.M. Hornos, and R.d.J. Napolitano.
	\newblock Continuously decoupling single-qubit operations from a perturbing
	thermal bath of scalar bosons.
	\newblock {\em Phys. Rev. A}, 75(2):022329, 2007.
	
	\bibitem{gordon2008optimal}
	G.~Gordon, G.~Kurizki, and D.~A. Lidar.
	\newblock Optimal dynamical decoherence control of a qubit.
	\newblock {\em Phys. Rev. Lett.}, 101(1):010403, 2008.
	
	\bibitem{Timoney2011}
	N.~Timoney, I.~Baumgart, M.~Johanning, A.~F. Varon, M.~B. Plenio, A.~Retzker,
	and C.~Wunderlich.
	\newblock Quantum gates and memory using microwave-dressed states.
	\newblock {\em Nature}, 476(7359):185--188, 2011.
	
	\bibitem{webster2013simple}
	S.C. Webster, S.~Weidt, K.~Lake, J.J. McLoughlin, and W.K. Hensinger.
	\newblock Simple manipulation of a microwave dressed-state ion qubit.
	\newblock {\em Phys. Rev. Lett.}, 111(14):140501, 2013.
	
	\bibitem{aharon2013general}
	N.~Aharon, M.~Drewsen, and A.~Retzker.
	\newblock General scheme for the construction of a protected qubit subspace.
	\newblock {\em Phys. Rev. Lett.}, 111(23):230507, 2013.
	
	\bibitem{wood2017microwave}
	J.D.A. Wood, J.~Tetienne, D.A. Broadway, L.T. Hall, D.A. Simpson, A.~Stacey,
	and L.C.L. Hollenberg.
	\newblock Microwave-free nuclear magnetic resonance at molecular scales.
	\newblock {\em Nat. Commun.}, 8, 2017.
	
	\bibitem{PhysRev.58.1098}
	T.~Holstein and H.~Primakoff.
	\newblock Field dependence of the intrinsic domain magnetization of a
	ferromagnet.
	\newblock {\em Phys. Rev.}, 58:1098--1113, 1940.
	
	\bibitem{SupplementalMaterial}
	Supplementary material.
	
	\bibitem{staudacher2015probing}
	T.~Staudacher, N.~Raatz, S.~Pezzagna, J.~Meijer, F.~Reinhard, C.A. Meriles, and
	J.~Wrachtrup.
	\newblock Probing molecular dynamics at the nanoscale via an individual
	paramagnetic centre.
	\newblock {\em Nat. Commun.}, 6:8527, 2015.
	
	\bibitem{PhysRevApplied.4.024004}
	X.~Kong, A.~Stark, J.~Du, L.P. McGuinness, and F.~Jelezko.
	\newblock Towards chemical structure resolution with nanoscale nuclear magnetic
	resonance spectroscopy.
	\newblock {\em Phys. Rev. Applied}, 4:024004, 2015.
	
	\bibitem{london2013detecting}
	P.~London, J.~Scheuer, J.-M. Cai, I.~Schwarz, A.~Retzker, M.~B. Plenio,
	M.~Katagiri, T.~Teraji, S.~Koizumi, J.~Isoya, R.~Fischer, L.~P. McGuinness,
	B.~Naydenov, and F.~Jelezko.
	\newblock Detecting and polarizing nuclear spins with double resonance on a
	single electron spin.
	\newblock {\em Phys. Rev. Lett.}, 111:067601, 2013.
	
	\bibitem{PhysRevB.75.155324}
	H.~Christ, J.~I. Cirac, and G.~Giedke.
	\newblock Quantum description of nuclear spin cooling in a quantum dot.
	\newblock {\em Phys. Rev. B}, 75:155324, 2007.
	
	\bibitem{plenio2004dynamics}
	M.B. Plenio, J.~Hartley, and J.~Eisert.
	\newblock Dynamics and manipulation of entanglement in coupled harmonic systems
	with many degrees of freedom.
	\newblock {\em New J. Phys.}, 6:36, 2004.
	
	\bibitem{eisert2003introduction}
	J.~Eisert and M.B. Plenio.
	\newblock Introduction to the basics of entanglement theory in
	continuous-variable systems.
	\newblock {\em Int. J. Quantum Inf.}, 1:479--506, 2003.
	
	\bibitem{Staudacher561}
	T.~Staudacher, F.~Shi, S.~Pezzagna, J.~Meijer, J.~Du, C.~A. Meriles,
	F.~Reinhard, and J.~Wrachtrup.
	\newblock Nuclear magnetic resonance spectroscopy on a (5-nanometer)$^3$ sample
	volume.
	\newblock {\em Science}, 339(6119):561--563, 2013.
	
	\bibitem{gardiner1985handbook}
	C.~W. Gardiner.
	\newblock {\em Handbook of stochastic methods}, volume~3.
	\newblock Springer Berlin, 1985.
	
	\bibitem{PhysRevB.93.045425}
	L.~M. Pham, S.~J. DeVience, F.~Casola, I.~Lovchinsky, A.~O. Sushkov, E.~Bersin,
	J.~Lee, E.~Urbach, P.~Cappellaro, H.~Park, A.~Yacoby, M.~Lukin, and R.~L.
	Walsworth.
	\newblock {NMR} technique for determining the depth of shallow nitrogen-vacancy
	centers in diamond.
	\newblock {\em Phys. Rev. B}, 93:045425, 2016.
	
	\bibitem{binder2017qudi}
	J.~M. Binder, A.~Stark, N.~Tomek, J.~Scheuer, F.~Frank, K.~D. Jahnke,
	C.~M{\"u}ller, S.~Schmitt, M.~H. Metsch, T.~Unden, T.~Gehring, A.~Huck, U.L.
	Andersen, L.~J. Rogers, and F.~Jelezko.
	\newblock {Qudi: A modular python suite for experiment control and data
		processing}.
	\newblock {\em SoftwareX}, 6:85--90, 2017.
	
	\bibitem{slichter1990principles}
	C.~P. Slichter.
	\newblock {\em Principles of Magnetic Resonance}, volume~1.
	\newblock Springer Science \& Business Media, 1990.
	
	\bibitem{meriles2010imaging}
	C.~A. Meriles, L.~Jiang, G.~Goldstein, J.~S. Hodges, J.~Maze, M.~D. Lukin, and
	P.~Cappellaro.
	\newblock Imaging mesoscopic nuclear spin noise with a diamond magnetometer.
	\newblock {\em J. Chem. Phys.}, 133(12):124105, 2010.
	
	\bibitem{Fullerton1982209}
	G.D. Fullerton, J.L. Potter, and N.C. Dornbluth.
	\newblock {NMR relaxation of protons in tissues and other macromolecular water
		solutions}.
	\newblock {\em J. Magn. Reson. Imaging.}, 1(4):209--226, 1982.
	
	\bibitem{Shapiro20111184}
	Y.~E. Shapiro.
	\newblock {Structure and dynamics of hydrogels and organogels: An NMR
		spectroscopy approach}.
	\newblock {\em Prog. Polym. Sci.}, 36(9):1184--1253, 2011.
	
\end{thebibliography}


\end{document}